\documentclass[conference]{IEEEtran}
\usepackage{blindtext, graphicx}
\usepackage[caption=false]{subfig}
\usepackage{booktabs,makecell}
\usepackage{amsmath}
\usepackage[capitalise]{cleveref}
\usepackage{color}
\usepackage{cite}
\usepackage[binary-units]{siunitx}
\usepackage{threeparttable}
\usepackage{multirow}
\usepackage{tikz}
\usepackage{tikzscale}
\usepackage{listings}  
\usepackage{tablefootnote}
\usepackage{float}
\usepackage{tabularx}
\usepackage{threeparttable}

\usetikzlibrary{shapes.misc}
\usetikzlibrary{arrows}

\newfloat{lstfloat}{htbp}{lop}
\floatname{lstfloat}{Listing}

\newcommand\blfootnote[1]{
  \begingroup
  \renewcommand\thefootnote{}\footnote{#1}
  \addtocounter{footnote}{-1}
  \endgroup
}

\newcommand{\secref}[1]{Sec.~\ref{#1}}
\newcommand{\figref}[1]{Fig.~\ref{#1}}
\newcommand{\tblref}[1]{Tbl.~\ref{#1}}

\newcommand{\x}{$\times$}
\newcommand{\Nrf}{N_\mathrm{RF}}
\newcommand{\Nout}{N_\mathrm{out}}
\newcommand{\uW}{\si{\micro\watt}}
\newcommand{\sqmm}{\si{\square\milli\meter}}

\begin{document}

\title{Design Automation for Binarized Neural Networks:\\A Quantum Leap Opportunity?}

\author{
	\IEEEauthorblockN{
	    Manuele Rusci\IEEEauthorrefmark{1},
		Lukas Cavigelli\IEEEauthorrefmark{2},
		Luca Benini\IEEEauthorrefmark{2}\IEEEauthorrefmark{1}
	}
	\IEEEauthorblockA{
	    \IEEEauthorrefmark{1} Energy-Efficient Embedded Systems Laboratory, University of Bologna, Italy -- manuele.rusci@unibo.it
	}
	\IEEEauthorblockA{
		\IEEEauthorrefmark{2}Integrated Systems Laboratory, ETH Zurich, Switzerland -- \{cavigelli, benini\}@iis.ee.ethz.ch
	}
}

\maketitle
\bstctlcite{IEEEexample:BSTcontrol}

\begin{abstract}
Design automation in general, and in particular logic synthesis, can play a key role in enabling the design of application-specific Binarized Neural Networks (BNN). This paper presents the hardware design and synthesis of a purely combinational BNN for ultra-low power near-sensor processing. 
We leverage the major opportunities raised by BNN models, which consist mostly of logical bit-wise operations and integer counting and comparisons, for pushing ultra-low power deep learning circuits close to the sensor and coupling it with binarized mixed-signal image sensor data. We analyze area, power and energy metrics of BNNs synthesized as combinational networks. Our synthesis results in GlobalFoundries 22\,nm SOI technology shows a silicon area of 2.61\,\sqmm{} for implementing a combinational BNN with 32\x 32 binary input sensor receptive field and weight parameters fixed at design time. This is 2.2\x{} smaller than a synthesized network with re-configurable parameters. With respect to other comparable techniques for deep learning near-sensor processing,  our approach features a 10\x{} higher energy efficiency.
\blfootnote{This project was supported in part by the EU's H2020 programme under grant no. 732631 (OPRECOMP) and by the Swiss National Science Foundation under grant 162524 (MicroLearn).}
\end{abstract}

\section{Introduction}
Bringing intelligence close to the sensors is an effective strategy to meet the energy requirement of battery-powered devices for always-ON applications~\cite{Alioto2017}. 
Power-optimized solutions for near-sensor processing aim at reducing the amount of data to be dispatched out from the sensors. Local data analysis can compress the data down to even a single bit in case of a binary classifier, hence massively reducing the output bandwidth and energy consumption over raw sensor data communication~\cite{Rusci2016}. 

In the context of visual sensing, novel computer vision chips feature embedded processing capabilities to reduce the overall energy consumption~\cite{Rodriguez2017}. By placing computational modules within the sensor, mid- and low- level visual features can be directly extracted and  transferred to a processing unit for further computation or used to feed a first stage classifier. 
Moreover, by integrating analog processing circuits on the focal-plane, 
the amount of data crossing the costly analog-to-digital border is reduced~\cite{Likamwa2016}. 
If compared with a camera-based system featuring a traditional imaging technology, this approach has a lower energy consumption because of (a) a reduced sensor-to-processor bandwidth and (b) a lower demand for digital computation~\cite{Zhang2016}. 
Relevant examples of mixed-signal smart capabilities include the 
extraction of spatial and temporal features, such as edges or frame-difference maps, or a combination of them~\cite{Fernandez2014}. Because of the employed highly optimized architectures, the power consumption of smart visual chips results to be more than one order of magnitude lower than off-the-shelf traditional image sensors~\cite{Gottardi2009}.

However, to favor the meeting between smart ultra-low power sensing and deep learning, which is nowadays the leading technique for data analytics, a further step is required. 
At present, the high computational and memory requirement of deep learning inference models have prevented a full integration of these approaches close to the sensor at an ultra low power cost~\cite{Likamwa2016,andri2017yodann}.
A big opportunity for pushing deep learning into low-power sensing come from recently proposed Binarized Neural Networks (BNNs)~\cite{Courbariaux2016,Rastegari2016}. When looking at the inference task, a BNN consists of logical XNOR operations, binary popcounts and integer thresholding. Therefore, major opportunities arise for hardware implementation of these models as part of the smart sensing pipeline~\cite{Umuroglu2017}.

In this paper, we explore the feasibility of deploying BNNs as a front-end for an ultra-low power smart vision chip. The combination of mixed-signal processing and hardware BNN implementation represents an extremely energy-efficient and powerful solution for always-ON sensing, serving as an early detector of interesting events. Therefore, we design and synthesize a purely combinational hard-wired BNN, which is fed with the binary data produced by a mixed-signal ultra-low power imager~\cite{Gottardi2009}.
The main contributions of this paper are:
\begin{itemize}
    \item The hardware design and logic synthesis of a combinational BNN architecture for always-ON near-sensor processing.
    \item The area and energy evaluation of the proposed approach, for varying network models and configurations.
\end{itemize}
We evaluate two BNN models with 16\x 16 and 32\x 32 binary input size, either with fixed or variable parameters. 
In case of a combinational BNN with 32\x 32 input data and hardwired parameters,
our synthesis results in GlobalFoundries 22\,nm SOI technology shows an area occupancy of 2.61\,\sqmm{}, which is 2.2\x{} smaller than the model with variable parameters, 
and features a 10\x{} higher energy efficiency with respect to comparable techniques for deep learning-based near-sensor processing. Moreover, our study paves the way for exploring a new generation of logic synthesis tools---aimed at aggressively optimizing deep binarized networks and enabling focal-plane processing of images with higher resolution. 

\section{Related Work}
Several proposed smart imaging chips for always-ON applications embed mixed-signal processing circuits for extracting basic spatial and temporal features directly on the sensor die~\cite{Choi2014,Fernandez2014,Kim2013}. Recent approaches tried to push deep learning circuits to the analog sensor side to exploit the benefits of focal-plane processing~\cite{Rodriguez2017}. The work presented in~\cite{Chen2016} makes use of angle-sensitive pixels, integrating diffraction gratings on the focal plane. Based on the different orientations of the pixel-level filters, multiple feature maps are locally computed as the first layer of a convolutional network. 
A sensing front-end supporting analog multiplication is proposed in~\cite{Lee2017}. They introduce a MAC unit composed of only passive switches and capacitors to realize a switched-capacitor matrix multiplier, which achieves an energy efficiency of 8.7\,TOp/s/W when running convolution operations.
RedEye~\cite{Likamwa2016} embeds column-wise processing pipelines in the analog domain to perform 3D convolutions before of the digital conversion. 
The chip is implemented in 0.18\,$\mu$m technology and needs 1.4\,mJ to process the initial 5 layers of GoogLeNet, leading to an energy efficiency of less than 2\,TOp/s/W.
With respect to these focal-plane analog approaches, we leverage the potentiality of BNNs to deploy a digital and optimized purely combinational network to notably increase the energy efficiency of near-sensor processing circuits. 

Many neural network accelerators have been reported in the literature, most of them with an energy efficiency in the range of few TOp/s/W~\cite{du2015shidiannao,cavigelli2017origami,sze2017efficient}. Several recent approaches have focused on quantizing the weights down to binarization in order to gain a significant advantage in memory usage and energy efficiency~\cite{sze2017efficient,andri2017yodann}, pushing it up to around 60\,TOp/s/W while advances in training methods have achieved accuracy losses of less than 1\% for this setup. A new approach has been to quantize also the activations down to binary with initial accuracy losses of up to 30\% on the ILSVRC dataset, these have improved to around 11\% over the last two years and even less for smaller networks on datasets such as CIFAR-10 and SVHN~\cite{Courbariaux2016,Rastegari2016,sze2017efficient}. During this time, some VLSI implementations have been published, most of them targeting FPGAs such as the FINN framework~\cite{Umuroglu2017,intel2016}. Only few ASIC implementations exist~\cite{intel2016,BRein2017,xnorpop2017}, of which XNOR-POP uses in-memory processing and reports the highest energy efficiency of 21.1\,TOp/s/W and thus less than the best binary-weight-only implementation. 

\section{Combinational Hardware BNN Design}
BNNs feature a single-bit precision for both the weights and the activation layers when performing inference. This makes the approach promising for resource-constrained devices, also considering the intrinsic 32$\times$ memory footprint reduction with respect to baseline full-precision models.
When applying the binarization scheme to a Convolutional Neural Network (CNN), a BNN features a stacked architecture of binary convolutional layers. Every layer transforms the $IF$ binary input feature maps 
into the $OF$ binary output feature maps through the well-known convolution operation. 
Because of the binary domain \{0,1\} of both the input data and the weight filters, the convolution kernel can be rewritten as
\begin{equation} \label{eq:conv_bin}
    \varphi(m,x,y) = \mathrm{popcount} (\textit{weights(m)}\;\mathrm{xnor}\; \textit{recField(x,y)}),
\end{equation}
where $\varphi(m,x,y)$ is the result of the convolution, $\textit{weights(m)}$ is the array of binary filter weights and $\textit{recField(x,y)}$ is the receptive field of the output neuron located at position $(x,y)$ of the $m$-th output feature map. 
The $\mathrm{popcount}(\cdot)$ function returns the numbers of asserted bits of the argument. 
Note that the convolution output $\varphi(m,x,y)$ is an integer value. As presented by~\cite{Courbariaux2016}, the popcount result is binarized after a batch normalization layer. 
However, the normalization operation can be reduced to a comparison with an integer threshold,
\begin{figure}
    \centering
    \includegraphics[width=1\linewidth]{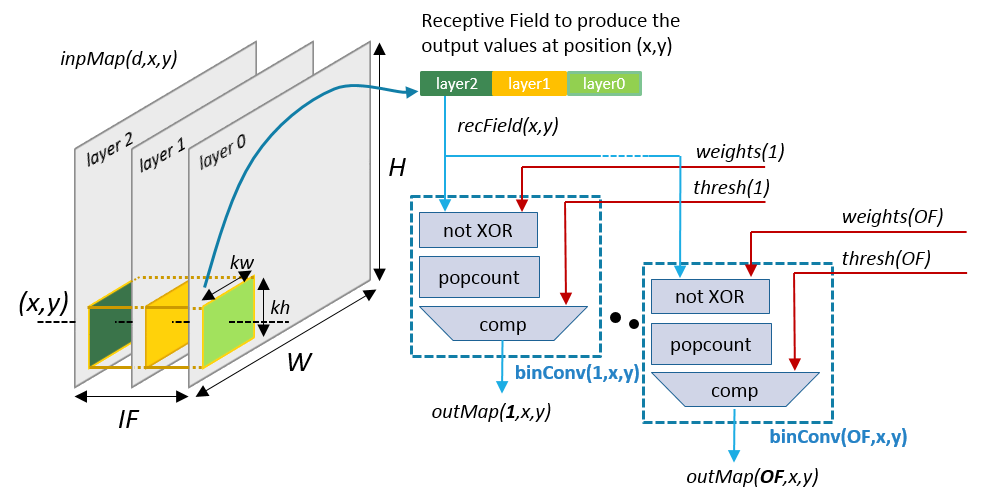}
    \caption{Binary convolution flow for every convolutional layer. For any of the OF output feature maps, the binary value at position $(x,y)$ is produced by overlapping the $m$-th weight filter to the array of the receptive field of the input feature map centered at the spatial position $(x,y)$.}
    \label{fig:BinaryConv}
\end{figure}
\begin{equation} \label{eq:bin_opt}
   outMap(m,x,y) =   \begin{cases}
    \varphi(m,x,y) \ge thresh(m)       &  \text{if } \gamma > 0 \\
    \varphi(m,x,y)  \le thresh(m)      & \text{if } \gamma < 0 \\
    1                         \qquad         \text{if } \gamma = 0 \text{ and } \beta \ge 0\\
    0                         \qquad         \text{if } \gamma = 0 \text{ and } \beta < 0
  \end{cases},
\end{equation}
where $thresh(m)$ is the integer threshold that depends on the convolution bias $b$ and on the parameters learned by the batch normalization layer $\mu$, $\gamma$, $\sigma$ and $\beta$. After training the network, the $thresh(m)$ parameters are computed offline as $\lfloor \mu - b - \beta \cdot \sigma/ \gamma \rfloor$ if $ \gamma > 0$ or $\lceil \mu - b - \beta \cdot \sigma/ \gamma \rceil$ if $\gamma < 0$.

\begin{figure}
    \centering
    \includegraphics[width=0.93\linewidth]{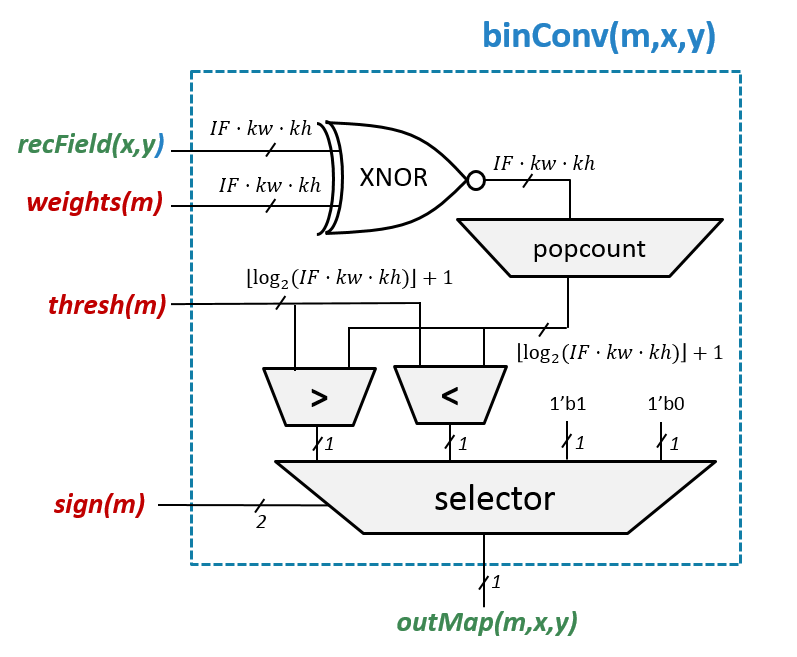}
    \caption{Hardware architecture of the combinational building block for computing binary convolutions. Every \textit{binConv(m,x,y)} module instantiated within a convolutional layer produces the binary value of the output neuron at location $(x,y)$ of the $m$-th output feature map.}
    \label{fig:BinaryConvBlock}
\end{figure}

\figref{fig:BinaryConv} graphically schematizes the binary convolution kernel. The \textit{BinConv} module applies \eqref{eq:conv_bin} and \eqref{eq:bin_opt} over the receptive field values of the output neuron $outMap(m,x,y)$. 
To build a convolutional layer, the \textit{BinConv} is replicated for every output neuron.
The hardware architecture of a \textit{BinConv} element is shown in \figref{fig:BinaryConvBlock}.
The input signals $\textit{recField(x,y)}$, $\textit{weights(m)}$ and $\textit{thresh(m)}$ and the output signal $\textit{outMap(m,x,y)}$ of the block 
refer to \eqref{eq:conv_bin} and \eqref{eq:bin_opt}. Additionally, the $\textit{sign(m)}$ signal drives the selection of the correct output neuron's value depending on the batch normalization parameters (eq. \eqref{eq:bin_opt}). The network parameters, \emph{weights, thresh} and \emph{sign}, highlighted in red, can be stored in a memory block, to allow online reconfiguration, or can be fixed at design time. In total, the memory footprint required to store the parameters of a convolutional layer is $OF \cdot (IF\cdot kw \cdot kh + \lfloor \log_2(IF\cdot kw \cdot kh) \rfloor + 3 )$ bits. 

Despite the reduced reconfigurability, relevant benefits in terms of silicon occupation arise when hard-wiring the binary weights. In this case, the synthesis tool plays a major role to enable the implementability of the model. The synthesis tool has to exploit the optimizations based on a high-level abstract HDL description of the network.

To explore the feasibility of deep combinational BNNs, we focus on VGG-like network topologies as in~\cite{Courbariaux2016}. These networks include convolutional layers with a small filter size (typically $kw=kh=3$) and an increasing feature dimension going deeper into the network. The spatial dimension tends to decrease by means of strided pooling operations placed after the binary convolution of \eqref{eq:conv_bin}. Following the intuition of~\cite{Umuroglu2017}, a MaxPooling layer can be moved behind the binarization by replacing the MAX with an OR operation among the binary values passing through the pooling filter. 

The VGG-like topology features multiple fully-connected layers. 
Their hardware implementation 
is similar to the \textit{binConv} module of \figref{fig:BinaryConvBlock}, where the convolutional receptive field contains all the input neurons of the layer. 
The last fully-connected layer generates a confidence score for every class.
Differently from the original BNN scheme, our network architecture is fed with a binary single-layer signal coming from a mixed-signal imager~\cite{Gottardi2009}. However, the presented approach also holds for multi-channel imagers.

\subsection{Estimating Area} \label{sec:areaEst}
Before looking at synthesis results, we estimate the area of a binary convolutional layer. 
For each output value (output pixel and feature map, $\Nout=H\cdot W\cdot OF$), we have a receptive field of size $\Nrf=IF\cdot kw\cdot kh$ and thus need a total of $\Nout\Nrf$ XNOR gates. These are followed by popcount units---adder trees summing over all $\Nrf$ values in the receptive field. 
The resulting full-precision adder trees require $\sum_{i=1}^{\log_2(\Nrf)} \Nrf 2^{-i}=\Nrf-1$ half-adders and $\sum_{i=1}^{\log_2(\Nrf)} (i-1) \Nrf 2^{-i}=\Nrf-\log_2(\Nrf)-1$ full-adders each, and are replicated for every output value. The subsequent threshold/compare unit is insignificant for the total area. 

To provide an example, we look at the first layer of the network for $16\times 16$ pixel images with 1 input and 16 output feature maps and a $3\times 3$ filter ($\Nrf=9, \Nout=4096$). 
Evaluating this for the GF22 technology with $A_\mathrm{XNOR}=0.73\,\mu\mathrm{m}^2$,  $A_\mathrm{HA}=1.06\,\mu\mathrm{m}^2$ and $A_\mathrm{FA}=1.60\,\mu\mathrm{m}^2$, we obtain an area of $A_\mathrm{XNOR,tot}=0.027\,\mathrm{mm}^2$, $A_\mathrm{HA,tot}=0.033\,\mathrm{mm}^2$ and $A_\mathrm{FA,tot}=0.029\,\mathrm{mm}^2$---a total of 0.089\,mm${}^2$. 
Note that this implies that the area scales faster than linearly with respect to the size of the receptive field $\Nrf$ since the word width in the adder tree increases rapidly. This is not accounted for in the widely used GOp/img complexity measure for NNs, as it is becoming only an issue in this very low word-width regime.

\section{Experimental Results}
\begin{table}
\caption{VGG-like BNN Models\tablefootnote{bConvLyr3x3($x$,$y$) indicates a binary convolutional layer with a 3\x 3 filter, $x$ input and $y$ output feature maps, MaxP2x2 is a max pooling layer of size 2\x 2, bFcLyr($x$, $y$) is a binary fully connected layer with $x$ binary input $y$ binary output binary neurons.}} 
\label{tab:bnnmodel}
\centering
\resizebox{1\columnwidth}{!}{
    \begin{tabular}{cll}
    \toprule
    layer & Model with a 16\x 16 input map  &  Model with a 32\x 32 input map \\ 
    \midrule
    1 & bConvLyr3x3( 1,16)+MaxP2x2 & bConvLyr3x3( 1,16)+MaxP2x2\\ 
    2 & bConvLyr3x3(16,32)+MaxP2x2 & bConvLyr3x3(16,32)+MaxP2x2\\
    3 & bConvLyr3x3(32,48)+MaxP2x2 & bConvLyr3x3(32,48)+MaxP2x2\\
    4 & bFcLyr(192,64)             & bConvLyr3x3(48,64)+MaxP2x2\\
    5 & bFcLyr( 64, 4)             & bFcLyr(256,64)\\
    6 &                            & bFcLyr( 64, 4)\\
    \bottomrule
    \end{tabular}
}
\end{table}

\subsection{BNN Training}
The experimental analysis focuses on two VGG-like network topologies described in \tblref{tab:bnnmodel} to investigate also the impact of different input and network size. 
As a case-study, we trained the networks with labelled patches from the MIO-TCD dataset~\cite{MIO-TCD} belonging to one of the following classes: cars, pedestrians, cyclist and background. The images from the dataset are resized to fit the input dimension before applying a non-linear binarization, which simulates the mixed-signal preprocessing of the sensor~\cite{Gottardi2009}. By training the BNNs with ADAM over a training set of about 10ksamples/class (original images are augmented by random rotation), the classification accuracy against the test-set achieves 64.7\% in case of the model with 32\x 32 input data, while a 50\% is measured for the 16\x 16 model because of the smaller input size and network. 
Since this work focuses on hardware synthesis issues of BNN inference engines, we do not explore advanced training approaches for NNs with non-traditional input data, which have been discussed in the literature~\cite{Jayasuriya2016}.

\subsection{Synthesis Results}
We analyze both aforementioned networks for two configurations, with weights fixed at synthesis time and with variable weights (excl. storage, modeled as inputs). The fixed weights are taken from the aforementioned trained models.

\begin{table}
    \centering
    \caption{Synthesis and Power Results for Different Configurations}
    \label{tbl:synthOverview}
    \begin{threeparttable}
    \begin{tabularx}{\linewidth}{@{\hskip 1mm}l@{\hskip 2mm}c@{\hskip 2mm}r@{\hskip 2mm}r@{\hskip 2mm}r@{\hskip 2mm}r@{\hskip 3mm}r@{\hskip 2mm}r@{\hskip 2mm}r@{\hskip 1mm}}
        \toprule
          & &\multicolumn{2}{c}{------\,area\,------}&   \multicolumn{2}{c}{---\,time/img\,---} &   E/img &   leak. & E-eff. \\
         netw. & type &   [\sqmm]   & [MGE]\tnote{\dag} &   [ns] &   [FO4]\tnote{\ddag} &   [nJ] &   [\uW{}] & [TOp/J] \\
        \midrule 
         16\x 16 & var.  &    1.17  &   5.87 &  12.82 &   560 &    2.40 &    945 &  470.8 \\
         16\x 16 & fixed &    0.46  &   2.32 &  12.40 &   541 &    1.68 &    331 &  672.6 \\
         32\x 32 & var.  &    5.80  &  29.14 &  17.27 &   754 &   11.14 &   4810 &  479.4 \\
         32\x 32 & fixed &    2.61  &  13.13 &  21.02 &   918 &   11.67 &   1830 &  457.6 \\
        \bottomrule
    \end{tabularx}
    \begin{tablenotes}
        \item[\dag] Two-input NAND-gate size equivalent: $1\,\mathrm{GE}=0.199\,\mu\mathrm{m}^2$
        \item[\ddag] Fanout-4 delay: $1\,\mathrm{FO4}=22.89\,\mathrm{ps}$
    \end{tablenotes}
    \end{threeparttable}
\end{table}
We provide an overview of synthesis results for different configurations in \tblref{tbl:synthOverview}. We synthesized both networks listed in \tblref{tab:bnnmodel} in GlobalFoundries 22\,nm SOI technology with LVT cells in the typical case corner at 0.65\,V and 25$^{\circ}$C. The configuration with variable weights scales with the computational effort associated with the network (1.13\,MOp/img and 5.34\,MOp/img for the 16\x 16 and 32\x 32 networks) with 0.97 and 0.92\,MOp/cycle/\sqmm{}, respectively. 
The variable parameters/weights configuration does not include the storage of the parameters themselves, which would add \SI{1.60}{\square\micro\meter} (8.0\,GE) per FF 
which could be loaded through a scan-chain without additional logic cells (from some flash memory elsewhere on the device). Alternatively, non-volatile memory cells could be used to store them. The number of parameters is 33 and 65\,kbit and thus 0.05\,\sqmm{} (264\,kGE) and 0.10\,\sqmm{} (520\,kGE) for the 16\x 16 and 32\x 32 network, respectively. 

\begin{table}
    \centering
    \caption{Area Breakdown for the 16\x 16 Network} 
    \label{tbl:areaDetail2}
    \begin{tabular}{cr|rrr}
        \toprule 
                       & compute & area estim. & var. weights & fixed weights \\
         layer         & [kOp/img] &   [\sqmm]    &   area [\sqmm] &   area [\sqmm] \\
        \midrule
         1  &  74 ( 6.5\%) & 0.093 &     0.077 ( 6.6\%) &      0.008 ( 1.7\%) \\
         2  & 590 (52.2\%) & 0.971 &     0.647 (55.4\%) &      0.204 (44.3\%) \\
         3  & 442 (39.1\%) & 0.738 &     0.417 (35.8\%) &      0.241 (52.3\%) \\
         4  &  25 ( 2.2\%) & 0.041 &     0.026 ( 2.2\%) &      0.008 ( 1.7\%) \\
        \bottomrule
    \end{tabular}
\end{table}
Looking at the more detailed area breakdown in \tblref{tbl:areaDetail2}, we can see that there is a massive reduction when fixing the weights before synthesis. Clearly, this eliminates all the XNOR operations which become either an inverter or a wire, and even the inverter can now be shared among all units having this particular input value in their receptive field. However, based on the estimates described in \secref{sec:areaEst}, this cannot explain all the savings. Additional cells can be saved through the reuse of identical partial results, which not only can occur randomly but must occur frequently. For example, consider 16 parallel popcount units summing over 8 values each. We can split the value into 4 groups with 2 values each. Two binary values can generate $2^2=4$ output combinations. Since we have 16 units of which each will need one of the combinations, they will on average be reused 4 times. This is only possible with fixed weights, otherwise the values to reuse would have to be multiplexed, thereby loosing all the savings. 

Generally, we can observe that these already small networks for low-resolution images require a sizable amount of area, such that more advanced ad-hoc synthesis tools exploiting the sharing of weights and intermediate results are needed.

\subsection{Energy Efficiency Evaluations}
\begin{table}
    \centering
    \caption{Energy and Leakage Breakdown for the 16\x 16 Network}
    \label{tbl:powerDetail}
    \begin{tabular}{crrrr}
        \toprule
                     & \multicolumn{2}{c}{-------- var. weights --------} & \multicolumn{2}{c}{-------- fixed weights --------} \\
         layer       & energy/img [pJ] & leakage & energy/img [pJ] & leakage \\
        \midrule
         1 &    38 ( 1.6\%) &  68\,\uW  &    9 ( 0.5\%) &   8\,\uW \\
         2 &   806 (33.7\%) & 547\,\uW  &  478 (28.5\%) & 152\,\uW  \\
         3 &  1440 (60.2\%) & 310\,\uW  & 1037 (61.9\%) & 163\,\uW  \\
         4 &   107 ( 4.5\%) &  20\,\uW  &  151 ( 9.0\%) &   7\,\uW  \\
        \bottomrule
    \end{tabular}
\end{table}
We have performed post-synthesis power simulations using 100 randomly selected real images from the dataset as stimuli. 
The results are also reported in \tblref{tbl:synthOverview} while a detailed per-layer breakdown is shown in \tblref{tbl:powerDetail}.
We see that the model with 32\x 32 input has lower energy-efficiency and higher latency when fixing the weights, while the opposite is observed for the smaller model. We attribute this to the fact that synthesis is set to optimize for area and both, the critical path length and target power are unconstrained. 
These energy efficiency numbers are in the order of 10\x{} higher than those of the next competitor YodaNN~\cite{andri2017yodann}. However, they are fundamentally different in the sense that YodaNN (a) runs the more complex binary weight networks, (b) requires additional off-chip memory for the weights and intermediate results, (c) can run large networks with a fixed-size accelerator, and (d) is in an older technology but doing aggressive voltage scaling. Given these major differences, a more in-depth comparison would require a redesign of YodaNN in 22\,nm and re-tuning to the single-channel input architecture we are using for comparison. Nevertheless, is is clear that these combinational BNNs are by far more efficient. 

When heavily duty-cycling a device, leakage can become a problem. In this case, we see 945\,\uW{} and 331\,\uW{} of leakage power, which might be significant enough in case of low utilization to require mitigation through power-gating or using HVT cells. Generally, voltage scaling can also be applied, not only reducing leakage, but also active power dissipation. 
The throughput we observe in the range of 50\,Mframe/s is far in excess of what is meaningful for most applications. Thus aggressive voltage scaling, power gating and the reverse body biasing available in this FD-SOI technology should be optimally combined to reach the minimum energy point where leakage and dynamic power are equal while the supply is ON. 

We expect these values to be highly dependent on the input data, since energy is consumed only when values toggle. While a single pixel toggling at the input might affect many values later in the network, it has been shown that rather the opposite effect can be seen: changes at the input tend to vanish deeper into the network~\cite{cbinfer}. A purely combinational implementation fully leverages this and BNNs naturally have a threshold that keeps small changes from propagating and might thus perform even better for many real-world applications. 

\subsection{Scaling to Larger Networks}
Our results show an area requirement in the range of 2.05 to 2.46\,GE/Op and an average 1.9\,fJ/Op. Scaling this up to \SI{0.5}{\square\centi\meter}~(250\,MGE) of silicon and an energy consumption of only 210\,nJ/img, we could map networks of around 110\,MOp/img---this is already more than optimized high-quality ImageNet classification networks such as ShuffleNets require~\cite{zhang2017shufflenet}. 

\begin{figure}
    \centering
    \includegraphics[width=1\linewidth]{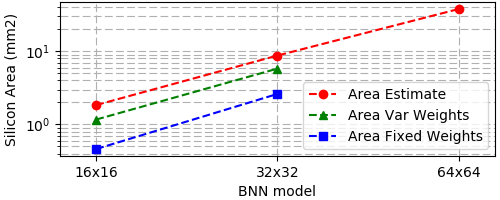}
    \caption{Silicon area estimation (in red) and measurements with variable (green) and fixed (blue) weights of three BNNs featuring a model complexity which scales depending on the imager resolution. The area occupation of the 64\x 64 model is not reported because the synthesis tool is not able to handle such a complex and large design.}
    \label{fig:plot}
\end{figure}

\figref{fig:plot} shows the estimation and measurements of the silicon area corresponding to the synthesized BNNs for fixed and variable weights. We also consider a model with a larger 64\x 64 input imager receptive field and a higher complexity (5 convolutional and 2 fully-connected layers, 23.05\,GOp/img). Such a model presents is more accurate on the considered classification task (73.6\%) but current synthesis tool cannot handle the high complexity of the design, using in excess of 256\,GB of memory. When estimating the area occupancy, the 64\x 64
BNNs result to be 4.3\x{} larger than the area estimated for the 32\x 32 model. 
A direct optimization of such large designs is out of scope of today's EDA tools, clearly showing the need for specialized design automation tools for BNNs. 

\section{Conclusion}
We have presented a purely combinational design and synthesis of BNNs for near-sensor processing. Our results demonstrate the suitability and the energy efficiency benefits of the proposed solution, fitting on a silicon area of 2.61\,\sqmm{} when considering a BNN model with 32\x 32 binary input data and weight parameters fixed at design time. Our study also highlighted the need for novel synthesis tools able to deal with very large and complex network designs, that are not easily handled by current tools.

\bibliographystyle{IEEEtran}
\bibliography{bstctl,refs}

\begin{thebibliography}{10}
\providecommand{\url}[1]{#1}
\csname url@samestyle\endcsname
\providecommand{\newblock}{\relax}
\providecommand{\bibinfo}[2]{#2}
\providecommand{\BIBentrySTDinterwordspacing}{\spaceskip=0pt\relax}
\providecommand{\BIBentryALTinterwordstretchfactor}{4}
\providecommand{\BIBentryALTinterwordspacing}{\spaceskip=\fontdimen2\font plus
\BIBentryALTinterwordstretchfactor\fontdimen3\font minus
  \fontdimen4\font\relax}
\providecommand{\BIBforeignlanguage}[2]{{%
\expandafter\ifx\csname l@#1\endcsname\relax
\typeout{** WARNING: IEEEtran.bst: No hyphenation pattern has been}%
\typeout{** loaded for the language `#1'. Using the pattern for}%
\typeout{** the default language instead.}%
\else
\language=\csname l@#1\endcsname
\fi
#2}}
\providecommand{\BIBdecl}{\relax}
\BIBdecl

\bibitem{Alioto2017}
M.~Alioto, \emph{Enabling the Internet of Things: From Integrated Circuits to
  Integrated Systems}.\hskip 1em plus 0.5em minus 0.4em\relax Springer, 2017.

\bibitem{Rusci2016}
M.~Rusci, D.~Rossi \emph{et~al.}, ``An event-driven ultra-low-power smart
  visual sensor,'' \emph{IEEE Sensors Journal}, vol.~16, no.~13, pp.
  5344--5353, 2016.

\bibitem{Rodriguez2017}
{\'A}.~Rodr{\'\i}guez-V{\'a}zquez, R.~Carmona-Gal{\'a}n \emph{et~al.}, ``In the
  quest of vision-sensors-on-chip: Pre-processing sensors for data reduction,''
  \emph{Electronic Imaging}, vol. 2017, no.~11, pp. 96--101, 2017.

\bibitem{Likamwa2016}
R.~LiKamWa, Y.~Hou \emph{et~al.}, ``Redeye: analog convnet image sensor
  architecture for continuous mobile vision,'' in \emph{Proc. IEEE ISCA}, 2016,
  pp. 255--266.

\bibitem{Zhang2016}
S.~Zhang, M.~Kang \emph{et~al.}, ``Reducing the energy cost of inference via
  in-sensor information processing,'' \emph{arXiv:1607.00667}, 2016.

\bibitem{Fernandez2014}
J.~Fern{\'a}ndez-Berni, R.~Carmona-Gal{\'a}n \emph{et~al.}, ``Focal-plane
  sensing-processing: A power-efficient approach for the implementation of
  privacy-aware networked visual sensors,'' \emph{Sensors}, vol.~14, no.~8, pp.
  15\,203--15\,226, 2014.

\bibitem{Gottardi2009}
M.~Gottardi, N.~Massari, and S.~A. Jawed, ``A 100$\mu$ w 128$\times$64 pixels
  contrast-based asynchronous binary vision sensor for sensor networks
  applications,'' \emph{IEEE Journal of Solid-State Circuits}, vol.~44, no.~5,
  pp. 1582--1592, 2009.

\bibitem{andri2017yodann}
R.~Andri, L.~Cavigelli \emph{et~al.}, ``Yodann: An architecture for ultra-low
  power binary-weight cnn acceleration,'' \emph{IEEE Transactions on
  Computer-Aided Design of Integrated Circuits and Systems}, 2017.

\bibitem{Courbariaux2016}
M.~Courbariaux, I.~Hubara \emph{et~al.}, ``Binarized neural networks: Training
  deep neural networks with weights and activations constrained to+ 1 or-1,''
  \emph{arXiv:1602.02830}, 2016.

\bibitem{Rastegari2016}
M.~Rastegari, V.~Ordonez \emph{et~al.}, ``Xnor-net: Imagenet classification
  using binary convolutional neural networks,'' in \emph{Proc. ECCV}.\hskip 1em
  plus 0.5em minus 0.4em\relax Springer, 2016, pp. 525--542.

\bibitem{Umuroglu2017}
Y.~Umuroglu, N.~J. Fraser \emph{et~al.}, ``Finn: A framework for fast, scalable
  binarized neural network inference,'' in \emph{Proc. ACM/SIGDA FPGA}, 2017,
  pp. 65--74.

\bibitem{Choi2014}
J.~Choi, S.~Park \emph{et~al.}, ``A 3.4-$\mu$w object-adaptive cmos image
  sensor with embedded feature extraction algorithm for motion-triggered
  object-of-interest imaging,'' \emph{IEEE Journal of Solid-State Circuits},
  vol.~49, no.~1, pp. 289--300, 2014.

\bibitem{Kim2013}
G.~Kim, M.~Barangi \emph{et~al.}, ``A 467nw cmos visual motion sensor with
  temporal averaging and pixel aggregation,'' in \emph{Proc. IEEE ISSCC}, 2013,
  pp. 480--481.

\bibitem{Chen2016}
H.~G. Chen, S.~Jayasuriya \emph{et~al.}, ``Asp vision: Optically computing the
  first layer of convolutional neural networks using angle sensitive pixels,''
  in \emph{Proc. IEEE CVPR}, 2016, pp. 903--912.

\bibitem{Lee2017}
E.~H. Lee and S.~S. Wong, ``Analysis and design of a passive switched-capacitor
  matrix multiplier for approximate computing,'' \emph{IEEE Journal of
  Solid-State Circuits}, vol.~52, no.~1, pp. 261--271, 2017.

\bibitem{du2015shidiannao}
Z.~Du, R.~Fasthuber \emph{et~al.}, ``Shidiannao: Shifting vision processing
  closer to the sensor,'' in \emph{ACM SIGARCH Computer Architecture News},
  vol.~43, no.~3, 2015, pp. 92--104.

\bibitem{cavigelli2017origami}
L.~Cavigelli and L.~Benini, ``Origami: A 803-gop/s/w convolutional network
  accelerator,'' \emph{IEEE Transactions on Circuits and Systems for Video
  Technology}, vol.~27, no.~11, pp. 2461--2475, 2017.

\bibitem{sze2017efficient}
V.~Sze, Y.-H. Chen \emph{et~al.}, ``Efficient processing of deep neural
  networks: A tutorial and survey,'' \emph{arXiv:1703.09039}, 2017.

\bibitem{intel2016}
E.~Nurvitadhi, D.~Sheffield \emph{et~al.}, ``Accelerating binarized neural
  networks: Comparison of fpga, cpu, gpu, and asic,'' in \emph{Proc. FPT},
  2016, pp. 77--84.

\bibitem{BRein2017}
K.~Ando, K.~Ueyoshi \emph{et~al.}, ``Brein memory: A 13-layer 4.2 k neuron/0.8
  m synapse binary/ternary reconfigurable in-memory deep neural network
  accelerator in 65 nm cmos,'' in \emph{Proc. VLSI Symposium}, 2017.

\bibitem{xnorpop2017}
L.~Jiang, M.~Kim \emph{et~al.}, ``Xnor-pop: A processing-in-memory architecture
  for binary convolutional neural networks in wide-io2 drams,'' in \emph{Proc.
  IEEE/ACM ISLPED}, 2017.

\bibitem{MIO-TCD}
\BIBentryALTinterwordspacing
``The traffic surveillance workshop and challenge 2017 (tswc- 2017),'' 2017,
  {MIO-TCD}: MIOvision Traffic Camera Dataset. [Online]. Available:
  \url{http://podoce.dinf.usherbrooke.ca}
\BIBentrySTDinterwordspacing

\bibitem{Jayasuriya2016}
S.~Jayasuriya, O.~Gallo \emph{et~al.}, ``Deep learning with energy-efficient
  binary gradient cameras,'' \emph{arXiv:1612.00986}, 2016.

\bibitem{cbinfer}
L.~Cavigelli, P.~Degen, and L.~Benini, ``Cbinfer: Change-based inference for
  convolutional neural networks on video data,'' \emph{arXiv:1704.04313}, 2017.

\bibitem{zhang2017shufflenet}
X.~Zhang, X.~Zhou \emph{et~al.}, ``Shufflenet: An extremely efficient
  convolutional neural network for mobile devices,'' \emph{arXiv:1707.01083},
  2017.

\end{thebibliography}

\end{document}